\documentclass[a4paper,aps,prb,twocolumn]{revtex4}
\usepackage{color}
\usepackage{array}
\usepackage{longtable} 
\usepackage{float}
\usepackage{graphicx}
\usepackage[unicode=true, pdfusetitle,  bookmarks=false,breaklinks=false,pdfborder={0 0 1},backref=false,colorlinks=false] {hyperref}

 

\begin{document}

\title{Simulation of reconstructions of the polar ZnO $(0001)$ surfaces}

\author{H. Meskine, P. A. Mulheran}

\affiliation{Department of Process \& Chemical Engineering, University Of Strathclyde, James Weir Building, 75 
Montrose, Glasgow G1 1XQ, United Kingdom}

\begin{abstract} Surface reconstructions on the polar ZnO(0001) surface are 
investigated using empirical potential models. Several possible reconstructions based around triangular motifs are 
investigated. The quenching of the dipole moment in the material dominates the energetics of the surface patterns so 
that no one particular size of surface triangular island or pit is strongly favoured. We employ Monte Carlo simulations 
to explore which patterns emerge from a high temperature quench and during deposition of additional ZnO monolayers. The 
simulations show that a range of triangular islands and pits evolve in competition with one another. The surface 
patterns we discover are qualitatively similar to those observed experimentally.
\end{abstract}
\maketitle

\section{Introduction}

Zinc oxide has a wide range of applicability from electronics to catalysis.\cite{ozgur_comprehensive_2005} For example 
it is used as part of ZnO/Me/ZnO (Me=metal) multilayer functional glass coatings designed to filter heat-generating 
infra-red solar radiation. This is usually achieved by incorporating a thin low emissivity metal layer a few nanometres 
thick.{\cite{Glasser_2000}} In the case where silver is used to construct a ZnO-Ag-ZnO sandwich, it has been 
shown\cite{barthel_asymmetric_2005,Ando1999308} that the lower Ag(111)/ZnO(0001) interface may fail for reasons not yet 
fully understood, leading to a sizable cost increase in the manufacturing of these devices. To understand this effect, 
one needs to characterise the interfacial structure that arises from the growth of Au on Zn(0001). A pre-requisite for 
this is a fundamental understanding of the Zn(0001) surface that templates this growth.

\begin{figure}[t]
\begin{centering}
\includegraphics[width=8cm]{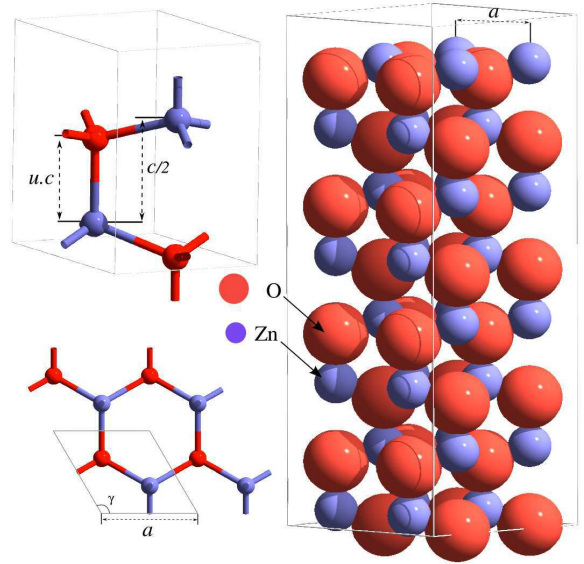}
\par
\end{centering}
\caption{Bulk crystal structure of wurtzite zinc-oxide with bulk lattice parameters $a=3.25$ \AA, $c=5.207$ \AA, and 
$u=0.3825$ \AA. In the bulk each ion is four-fold coordinated, while the surface atoms have only three-fold 
coordination. (colour online)\label{fig:01-Bulk-crystal-structure}}

\end{figure}

Zinc oxide (zincite) has the well-known wurtzite structure with lattice parameters at room temperature and ambient 
pressure of $a=3.25\mbox{ \AA}$, $c=5.207\mbox{ \AA}$, and $u=0.3825$, and space-group $P6_{3}mc$ (no. 186 in 
crystallographic tables).\cite{abrahams_remeasurement_1969,albertsson_atomic_1989} The structure may be understood as 
two interpenetrating hexagonal lattices, with each Zn (resp. O) sitting at the centre of a distorted O (resp. Zn) 
tetrahedron. The crystal when cut along the (0001) or (000$\overline{1}$) planes is known to be a type III polar 
material according to the Tasker classification. \cite{noguera_polar_2000,catlow_zinc_2008,tasker_stability_1979} That 
is to say that the unit cell is comprised of alternative negative and positive charged layers. This ultimately leads to 
a diverging electrostatic potential and should make the two polar surfaces of ZnO energetically unfavourable. This, 
however, is not the case as both the O-terminated and Zn-terminated polar surfaces show remarkable 
stability.\cite{noguera_rpp_2007}

Consider a slab of the material with bulk-terminated polar surface as used in typical computations (see Figure 1). 
Since the polar ZnO(0001) and ZnO(00$\bar{1}$) surfaces occur naturally, there must be a mechanism to quench the dipole 
moment that exists normal to the slab surface. In order to quench this macroscopic dipole moment a transfer of charge 
across the slab of $(1-2u)\sigma\approx0.235\times\sigma$ is necessary, where $\sigma$ is the surface charge 
density.\cite{noguera_rpp_2007} This may be understood in terms of the electrostatic energy change when charge is moved 
from one surface to the other in the direction of the internal electric field of the unreconstructed slab. Once 
sufficient charge has been moved, the counter electric field thus established cancels the one due to the bulk 
structure.

There are several mechanisms which may compensate the charge at the surface and counteract the macroscopic dipole 
moment of the semi-infinite crystal. Three, not necessarily incompatible, mechanisms have been considered in the 
literature: (\emph{i}) adsorption of charged species e.g. hydroxilation, (\emph{ii}) modification of the surface region 
by reconstruction, and (\emph{iii}) direct charge transfer.

Until recently, the exact nature of such a mechanism in ZnO was not well understood, but recent theoretical and 
experimental studies\cite{dulub_novel_2003,dulub_stm_2002,kresse_competing_2003} may have resolved the issue. A 
combination of surface microscopy techniques and Density Functional Theory (DFT) have indicated that, depending on the 
atmospheric environment, mechanisms (i) and (ii) may be active in quenching the polarisation of the ZnO(0001) surfaces. 
For the Zn terminated orientation and in hydrogen-rich conditions, the surface is best passivated by adsorption of 
hydroxyl groups, while under low hydrogen partial pressure the surface tends to form triangular reconstructions that 
appropriately compensate the charge imbalance created by the surface cut. This work seems to rule out the third 
mechanism (iii) which had previously been proposed,\cite{wander_stability_2001}involving charge transfer between 
O-terminated and Zn-terminated surfaces.

The above theoretical studies have relied on DFT which, while accurate, has a high computational cost and does not 
allow for a comprehensive search of the phase space. For example, in the case of the triangular reconstruction, STM 
scans show a range of triangle sizes while DFT studies only allow comparison of the energetics of single, relatively 
small configurations. In this work we present a study which combines fast, albeit less accurate, empirical potentials 
with Monte-Carlo simulations to study the structure and energetics of the polar ZnO(0001) surface, focusing on the 
surface reconstruction mechanism to quench the dipole. This approach is justified by the dominant role electrostatics 
plays in the surface resconstructions.\cite{kresse_competing_2003}

The rest of the paper has the following structure. Section II describes the methodology employed, discussing the 
empirical potentials, surface relaxation calculations and Monte Carlo (MC) simulations. The results are presented in 
Section III, firstly for the energetics of various surface reconstructions, and then for the surface patterns that 
emerge from the MC simulations. The implications of the results are discussed in the following section, and our 
conclusions are given in the final Section V.

\begin{figure}[t]
\begin{centering}
\includegraphics[width=8cm]{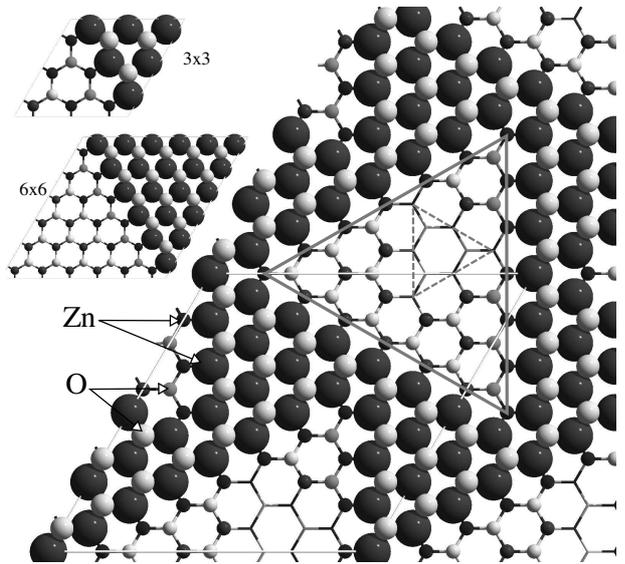}
\par
\end{centering}

\caption{Surface reconstruction $\sqrt{48}\times\sqrt{48}$ with a $n=7$ triangular pit and an additional $m=3$ inner 
pit within the larger triangle. The pits and terraces are created by removing Zn and O atoms. The topmost layer atoms 
are shown as large spheres (light red for O, dark blue for Zn) and the next layer atoms are shown by smaller spheres. 
The {}``bulk'' atoms in the slab (everything below layer 2) are shown only by their bonds. Upper left we show the 
smaller triangular pits in a $3\times3$ and $6\times6$ surface unit cell. \label{fig:02-initial-Triangle}}

\end{figure}

\section{Methodology}

\subsection{Empirical potentials and surface slab calculations}

Previous work by Catlow and coworkers \cite{catlow_zinc_2008,whitmore_surface_2002} has shown that empirical potentials 
are well suited to describe the details of the structure of the polar surfaces of most oxides. The parameters for a 
Buckingham potential are fitted following ref. \cite{binks_incorporation_1993} to reproduce a variety of properties of 
Zinc Oxide. The potential was kept as simple as possible as is appropriate for the desired level of computation. For 
details of the validation of the interatomic potential parameters we refer the reader to ref. 
\cite{woodley_database_2009}. The total energy is computed by summing all pair interactions of the form\begin{equation} 
E(r_{ij})=\frac{q_{i}q_{j}}{4\pi\epsilon_{0}r_{ij}}+A\exp(-r_{ij}/\rho)-C/r_{ij}^{6}\label{eq:E_rij}\end{equation} 
where $r_{ij}=\left\Vert \mathbf{r}_{i}-\mathbf{r}_{j}\right\Vert $ is the distance between two ions with charges 
$q_{i}$ and $q_{j}$. In this work we use formal ionic charges $\pm2e$ in all our computations. The first term of Eq. 
\ref{eq:E_rij} is the long-range Coulomb pair interaction, while the second and third terms correspond respectively to 
the repulsive and attractive terms of the short-range Buckingham pair potential.

The polarisability effects are described by a core-shell model, where the oxygen ion and its electronic cloud are 
modelled by a massive core and a mass-less shell carrying different charges (but with total charge $-$2e) and linked by 
a spring with energy\begin{equation} E^{\mbox{spring}}=\frac{1}{2}k_{i}\delta_{i}^{2},\label{eq:Espring}\end{equation} 
where $k_{i}$ is the spring constant for ion $i$ and $\delta_{i}$ is the core-shell distance. The empirical parameters 
$A$, $\rho$, $C$ and $k_{i}$ are determined by fitting to available experimental properties, such as the elastic 
constants. The fitting was performed with the GULP code\cite{gale_gulp:computer_1997} using 8 potential parameters and 
various parameters from observable data. The computed bulk properties are compared to some relevant experimental values 
in Table \ref{tab:pot-fit}, and the values of the parameters used in this work are given in Table 
\ref{tab:pot-param}.

\begin{table}[t]
\caption{Comparison of some of the computed bulk properties to available experimental data. The 
experimental data are taken from standard tables \cite{CRC}.
\label{tab:pot-fit}}

\noindent \centering{}
\begin{tabular}{ccc}
\hline
ZnO (wurtzite) & This work & Experiment\tabularnewline \hline 
$a(\mbox{\AA})$ & 3.27 & 3.25\tabularnewline $c(\mbox{\AA})$ & 5.18 & 5.207\tabularnewline $u$ & 0.3819 & 
0.3825\tabularnewline $\varepsilon_{11}^{T}$ & 4.22 & 9.26\tabularnewline $\varepsilon_{33}^{T}$ & 4.59 & 
11.0\tabularnewline $C_{11}$(GPa) & 222.22 & 209.7\tabularnewline $C_{33}$(GPa) & 220.14 & 210.9
\tabularnewline 
\end{tabular}
\end{table}

Surface structure calculations were performed using three-dimensional periodic slabs with a large vacuum gap normal to 
the (0001) surface. Each slab contains one Zn-terminated surface and one O-terminated surface (see Fig. 1). Several 
reconstructions were created at the surfaces, where overall charge neutrality was ensured by removing the oppositely 
charged species from the other side of the slab. The outermost three surface layers on each side of the slab were 
allowed to relax. The energy of different surface reconstructions were computed and an energy hierarchy constructed by 
comparing their surface energies. In order to estimate the surface energy of a given surface structure we use the total 
energy of the bulk unit cell as a reference state, in which case the surface energy is given by

\begin{equation} 
\gamma=\left(E_{slab}-N_{cell}\times\varepsilon_{bulk}\right)/2A\label{eq:SurfaceEnergyForm}
\end{equation}
where $A=\left(N_{s}a\right)^{2}\sin\gamma$ is the surface area of a slab (see Fig. \ref{fig:01-Bulk-crystal-structure}) 
containing $N_{s}\times N_{s}\times N_{z}$ lattice units, $E_{slab}$ is its relaxed total energy, and 
$\varepsilon_{bulk}$ is the energy of a bulk unit cell. The quantity $N_{cell}$ is the effective number of unit cells 
in the slab calculated by dividing the number of atoms in the slab by the number of atoms in a bulk unit cell ( four in 
the case of ZnO). The factor of two accounts for the fact that the surface energy is the average surface energy of both 
sides of the slab.

It is important to emphasise again that the surface energy calculated in this way depends on slab thickness, unless we 
have perfect quenching of the dipole moment by creating a net charge transfer of $0.235\times\sigma$ from one surface 
to the other in the reconstruction. For this reason, we will compare the energies of various surface structures using 
the same slab thickness $N_{z}=6$ (twelve bilayers).

\subsection{Monte-Carlo simulations}

While a large number of structures may be explored using the above empirical model, it is impractical to find the 
lowest energy reconstruction using more and more elaborate guesses of the surface structure. In order to explore the 
large phase space of possible surface reconstructions we used Monte Carlo (MC) simulations with bulk lattice positions 
in a slab. The ions in the three upper-most bilayers of the slab are allowed to hop within their own bilayer and into 
the bilayers directly above or below. The simulations were started from different initial configurations and the system 
left to evolve according to the Metropolis algorithm. The initial configurations of the three uppermost bilayers were 
formed either by ion removals or by addition of ions to the clean slab. The overall charge neutrality was again ensured 
by adding/removing the oppositely charged species from the other side of the slab at the start of each simulation. As 
several studies\cite{noguera_polar_2000} have shown that the surface layer relaxation is less than 0.1$\mbox{\AA}$, for 
the sake of simplicity (and computational efficiency) we have neglected the effect of lattice relaxation on surface 
energy during these MC simulations.

\begin{table}[t] 
\caption{
Interatomic potential parameters used for the Buckingham potential. These were obtained by fitting to the 
experimental parameters of Tab. The spring Constants are in eV$.$\AA $^{-2}$:  $k_{O}=15.52$ , $k_{Zn}=8.57$.
\label{tab:pot-param}
}
\noindent

\centering{}
\begin{tabular}{ccccc}
\hline
             & $A$(eV) & $\rho$(\AA) & $C$(eV$\times$\AA$^{6}$) & $r_{cutoff}$ \tabularnewline
\hline
Zn$-$O$^{s}$           & 499.6       &  0.359      &     0.0    &    0$-$10 \AA \tabularnewline
O$^{s}$$-$O$^{s}$      &  22764.0    &     0.149   & 27.88      &    0$-$12 \AA \tabularnewline
\hline 
\end{tabular}

\end{table}

A periodic slab model is used throughout, with $N_{z}=6$ bulk unit cells along the $c$ axis and a large vacuum of 
$L_{z}=30$\AA ~ added to form the super-cell. We verify that $L_{z}$ is large enough by ensuring that the surface 
energy does not depend on $L_{z}$.

Successive configurations are generated by a series of nearest neighbour hops of either species selected at random at 
the uppermost Zn-terminated surface, with acceptance probability $e^{-\Delta E/k_{B}T}$ where $T$ is the temperature 
and $\Delta E$ is the difference between total energies of the successive trial configurations. The bottom O-terminated 
surface reconstruction remains fixed in the simulations. The energy of a given atom in the slab is simply the sum of 
all its pair interaction with the other ions in the slab, with the pair interaction of ion $i$ given by

\begin{equation} 
\varepsilon_{i}=\sum_{j}E(r_{ij}).\label{eq:PairInteractionForm}
\end{equation}

Here, also for computational efficiency, 
we neglect the shell model component of the potential $E^{spring}$ in the MC work only.

Since in this system we have no mechanism to quench any dipole across the slab, we ensure that the simulation is 
started from configurations with only a small residual dipole. The initial configuration of the MC simulation is then 
disordered by running the simulation at very high temperature leading to a fully disordered arrangement of the surface 
species, after which the temperature is lowered abruptly. The simulations are run at high temperature for a large 
enough number of steps that the initial ordering disappears.

The most expensive step of the simulation is the energy evaluation which includes long-range terms. The Coulomb sum 
being conditionally convergent in a periodic system, we make use of the Ewald sum 
\cite{yeh_ewald_1999,brdka_a_ewald_2004,brdka_a._electrostatic_2002}
\begin{eqnarray}
E_{R} & = & 
\frac{1}{2}\sum_{\mathbf{n}}{}^{^{\prime}}\sum_{i,j}q_{i}q_{j}\frac{\mbox{erfc}(\alpha 
r_{ij,\mathbf{n}})}{r_{ij,\mathbf{n}}}\label{eq:EwaldReal}\\ E_{K} & = & 
\frac{2\pi}{V}\sum_{\mathbf{k}\ne\mathbf{0}}\frac{e^{-k^{2}/4\alpha^{2}}}{k^{2}}\left|S(\mathbf{k)}\right|^{2}\label{eq:EwaldRecip}\\ 
E_{0} & = & -\frac{\alpha}{\sqrt{\pi}}\sum_{i}q_{i}^{2}+\frac{2\pi}{V}M_{z}^{2}\label{eq:EwaldZero}
\end{eqnarray}
where 
$q_{i}$ is the formal charge of on $i$, $\mathbf{r}_{i}$is the position of the ion within the periodic slab, 
$S(\mathbf{k})=\sum_{i}q_{i}e^{i\mathbf{k}\mathbf{r}_{i}}$ is the structure factor, and $M_{z}$ is the $z$ coordinate 
of the total dipole moment in the slab $\mathbf{M}=\sum_{i}q_{i}\mathbf{r}_{i}$. The parameter $\alpha$ is determined 
using the requirement that the Ewald sum is accurate yet efficient (see for example \cite{fincham_optimisation_1994}). 
It is worth noting that this expression is the more computationally efficient 3D version of the Ewald sum, not the 
two-dimensional version. If the vacuum slab is chosen large enough, only a correction due to the residual surface 
dipole is necessary.

The above total energy is computed once at the beginning of the run, and updated in the course of the simulation by 
only computing the energy difference between trial configurations. This considerably speeds up the computation of the 
energy and scales as $N^{1/2}$, where $N$ is the number of particles in the system.

\section{Results}

\subsection{Surface reconstructions and energy hierarchy}

 \begin{table}[t]
 \centering{}
 \caption{Summary of the surface energies of several slabs, all with $N_{z}=6$, and 
various total number of ions $N$. For the relaxed structures. the total energies are computed using GULP with the 
potential parameters listed in table \ref{tab:pot-param}. The three outermost bilayers are allowed to relax while the 
rest of the slab is kept fixed. The vacancy concentration $\Theta_{\mbox{vac}}$ refers to the ratio of excess zinc 
\emph{vacancies} in the uppermost surface layer created by removing zincs and oxygen from both sides of the 
slab.
\label{tab:surface-energy}
}
\begin{longtable}{rcccc}
\hline
 & $\Theta_{\mbox{vac}}(\mbox{ML})$ & $N$ & \multicolumn{2}{c}{$\gamma(eV/\mbox{\AA}^{2})$}\tabularnewline
 & & & rigid & relaxed\tabularnewline \endhead \hline Bulk & - & - & - & \tabularnewline $2\times2$ & 0.500 & 88 & 
4.210 & 4.162\tabularnewline $4\times4$ & 0.250 & 352 & 0.255 & 0.086\tabularnewline $\sqrt{48}\times\sqrt{48}$ & 0.208 
& 1034 & 0.306 & 0.245\tabularnewline $6\times6$ & 0.055 & 856 & 2.248 & 1.447\tabularnewline
\end{longtable} 
\end{table}

\begin{figure}[t]
\begin{centering} 
\includegraphics[width=8cm]{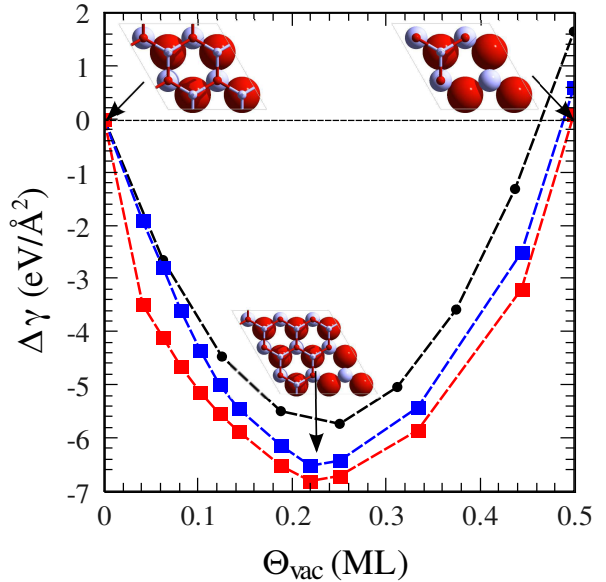} \par
\end{centering}

\caption{Change in surface energy as a function of excess zinc vacancy on the surface expressed in fractions of a 
monolayer, where $\Delta\gamma=\gamma_{nvac}-\gamma_{clean}$ is the net surface energy compared to the surface energy 
of the bulk-terminated surface. The blue line refers to bulk position of the atoms while the red curve correspond to 
optimised position for the three bilayers nearest the surface. The surface energy change when creating \emph{isolated} 
vacancies is shown by the black curve. Note that the small spheres refer to the topmost bilayer, while the larger ones 
refer to the bilayer immediately below it. \label{fig:03-Surface-Energy-Change}}

\end{figure}
On the zinc-terminated surface, it was experimentally shown that the triangular reconstructions are a single bilayer 
high. We therefore form the reconstructions by removing Zn and O atoms from the top-most bilayer and refer in what 
follows to the excess zinc vacancy concentration $\Theta_{vac}$. Recall that we remove the opposite charge species from 
the O-terminated surface, keeping overall charge neutrality and imposing an electric field across the slab which 
compensates the bulk dipole moment for $\Theta_{vac}=0.235$. The stability of the surface reconstruction is assessed by 
studying the surface energy as a function of this excess Zn vacancy concentration. Table \ref{tab:surface-energy} shows 
the surface energies for various zinc vacancy concentrations and their corresponding surface reconstructions 
illustrated in Figure \ref{fig:02-initial-Triangle} . Note that the reconstruction in the $\sqrt{48}\times\sqrt{48}$ 
surface unit cell as shown in Fig. \ref{fig:02-initial-Triangle} is not the most stable in this work, contrary to the 
prediction from density functional theory. This may be explained by a simple electrostatic argument, where the formal 
ionic representation of the species tends to overestimate the contribution of the Coulomb interaction, and thus favours 
smaller surface reconstructions.

We have computed the surface energies of several more surface reconstructions, for varying values of the excess zinc 
concentration, as well as for isolated vacancies. The results are summarised in Fig. \ref{fig:03-Surface-Energy-Change} 
and are plotted in reference to the surface energy of the bulk-terminated slab with $N_{z}=6$. We reiterate that the 
surface energy of the bulk-terminated slab is ill-defined due to the presence of a large surface dipole. It is only the 
\emph{difference} in surface energies of various reconstructions that is of interest. The most stable reconstruction is 
the small $4\times4$ cluster with the smallest triangular vacancy (see Figure \ref{fig:02-initial-Triangle}), with the 
more exotic reconstruction on the $\sqrt{48}\times\sqrt{48}$ surface cell lying nearby. The isolated vacancies 
consistently have a larger surface energy. Relaxation of the surface layers has a small effect on the surface 
stability, but does not impact the overall ordering of the various reconstruction energies.

In conclusion, the surface hierarchy obtained using the simple empirical potential shows no preference for large 
triangular reconstructions. Instead, small clusters which locally quench the surface dipole are preferred. The small 
impact of surface relaxation reflects the dominance of dipole moment quenching. This means that surface relaxation can 
be neglected in the MC simulations which follow, in which we allow the system to explore configuration space to see if 
larger reconstructions emerge naturally in our model with formal ionic charges.

\subsection{Monte-Carlo Simulations}

\begin{figure*}[t]
\begin{centering}
\includegraphics[height=12cm]{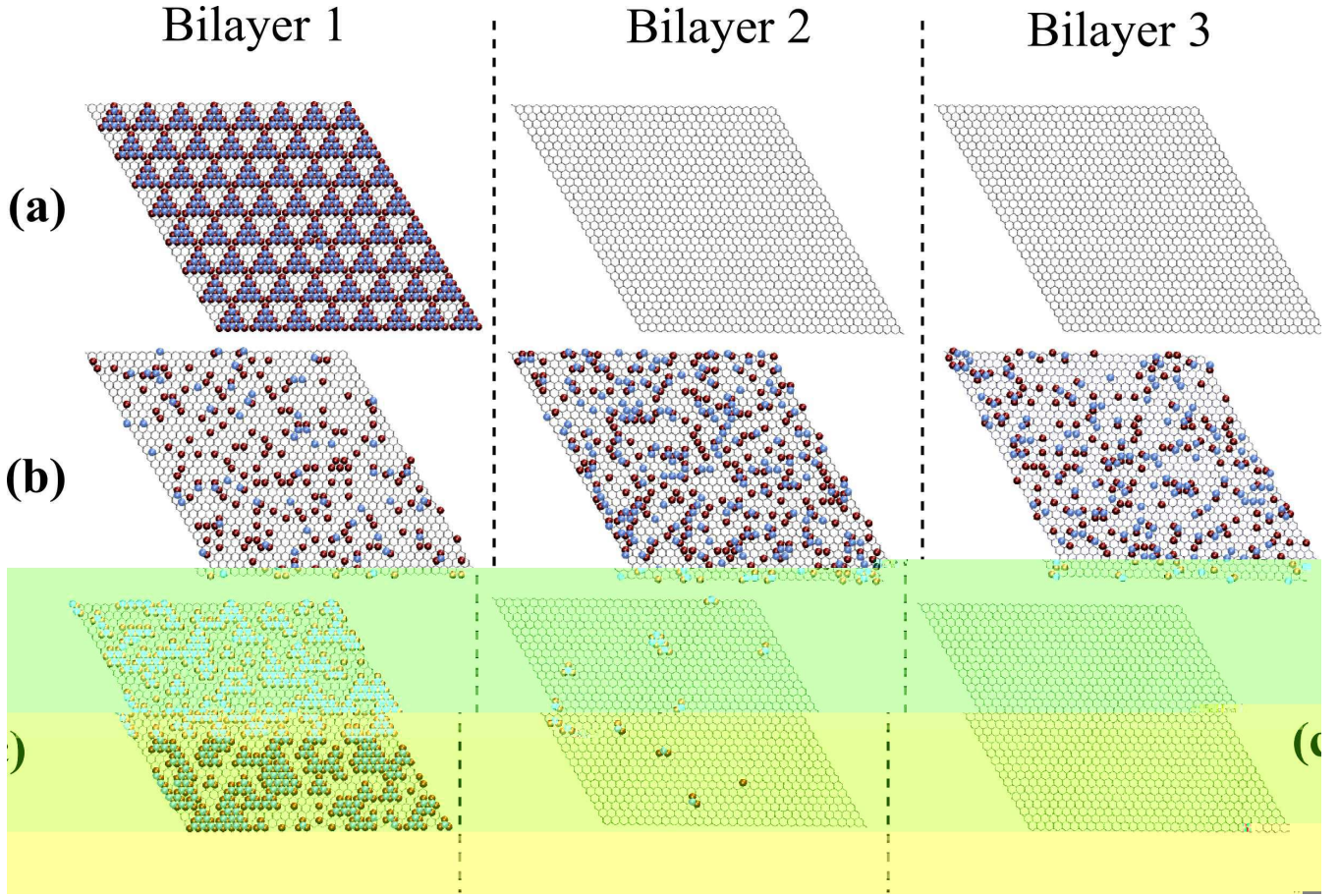} \par
\end{centering}

\caption{Snapshot of Zn$^{2+}$ and O$^{2-}$ ions in the three topmost layers at different stages of the evolution of 
the MC simulation. The initial configuration (a) is evolved at high temperature until the initial reconstructions are 
melted giving rise to a random distribution of species in all three layers (b). The temperature is then abruptly at 
which point triangular structures begin to spontaneously in bilayer 1, while the top bilayers are gradually 
emptied.(c).\label{fig:04-Snapshot-of-three bilayers}}

\end{figure*}

We have performed the MC simulation of a high temperature quench on a slab with $N_{z}=6,\, N_{s}=32$ and with the 
surfaces initially tessellated with the $4\times4$ triangular reconstructions giving $N=22,528$ ions in total. 
Representative snapshots of the species distribution in the three uppermost layers of the Zn-terminated surface are 
shown in Fig. \ref{fig:04-Snapshot-of-three bilayers}(\emph{a-c}). We show the three uppermost bilayers separately 
(bilayer 1-3) at \emph{(a)} the start of the simulation, \emph{(b)} at the end of the high temperature run, and 
\emph{(c)} after quenching. The figure clearly shows the effect of the high temperature, with the ordered triangular 
reconstructions completely disappearing from the first bilayer. The second and third bilayers are occupied as well, 
with no apparent ordering. After the system is cooled and left to evolve, large triangular reconstructions in the 
lowest bilayer begin to nucleate via aggregation of small triangular units. The smallest such unit is the one unit-cell 
triangle (three O ions surrounding one Zn ion) and is consistent with the earlier empirical potential result. The 
second bilayer only shows a few isolated $a-$side triangles, while the third layer is now completely empty. These 
observations are qualitatively identical for a large class of system size and parameters.

In Fig. 5 we show the evolution of the surface energy and the dipole moment normal to the surface in the simulation. 
During the high temperature anneal, both the surface energy and dipole moment are large in magnitude. This is due to 
the almost random placement of the surface layer ions into the 3 accessible bilayers, which leads to an obvious loss of 
bonding energy. The dipole moment also increases in magnitude since there is an excess of oxygen ions over zinc in the 
surface, so displacing them on average by one bilayer changes the total moment in the system. Upon quenching, the 
surface energy and dipole moment quickly reduce in magnitude, and in fact reach slightly lower values than in the 
starting configuration. At the start of the simulation, the dipole moment is $-162.29$ e$.$\AA, and is not zero since 
$\Theta_{vac}=0.25$ rather than the ideal $\Theta_{vac}=0.235$. After the temperature is raised, the dipole moment is 
roughly $-300$ e$.$\AA , since the topmost ions are now randomly distributed. By the end of the quench, after 
$N_{MC}=4\times10^{5}$, its value is slightly lower, $-236.15$ e$.$\AA . This is achieved by the few ions occupying the 
second bilayer. For the surface energy, the quenched value is 0.36 eV$/$\AA $^{2}$, as opposed to 0.59 eV$/$\AA $^{2}$ 
after heating.

\begin{figure}[t]
\noindent
\begin{centering}
\includegraphics[width=8cm]{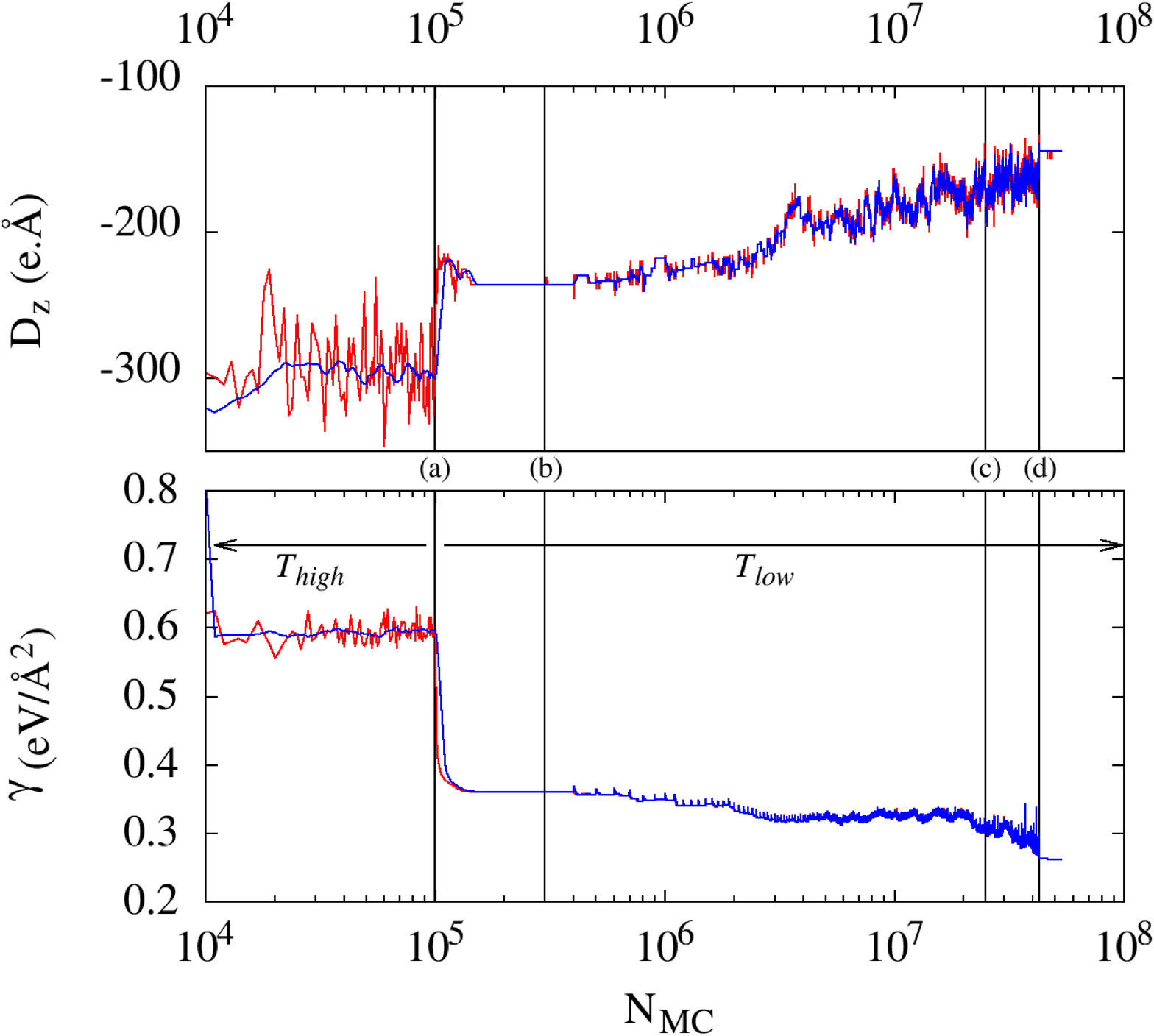}
\par
\end{centering}

\caption{Surface energy and dipole moment computed during a full Monte-Carlo run with , $N_{s}=32$. The simulation is 
started from a $4\times4$ tiled surface with a small residual surface dipole. We proceed to run the simulation at high 
temperature ($T_{high}$) for a large enough number of steps, then (a) quench to a lower temperature ($T_{low}$). The 
surface energy and dipole moment both settle at a lower value than initially. (b) Successive deposition of Zn/O pairs 
are then executed, up to 1 ML, then (c) 2 ML, allowing for a large enough number of steps between events to reach a new 
steady-state (d).\label{fig:05-surface-energy_dipole}}

\end{figure}

\begin{figure*}[t]
\noindent 

\begin{centering} 
\includegraphics[height=12cm]{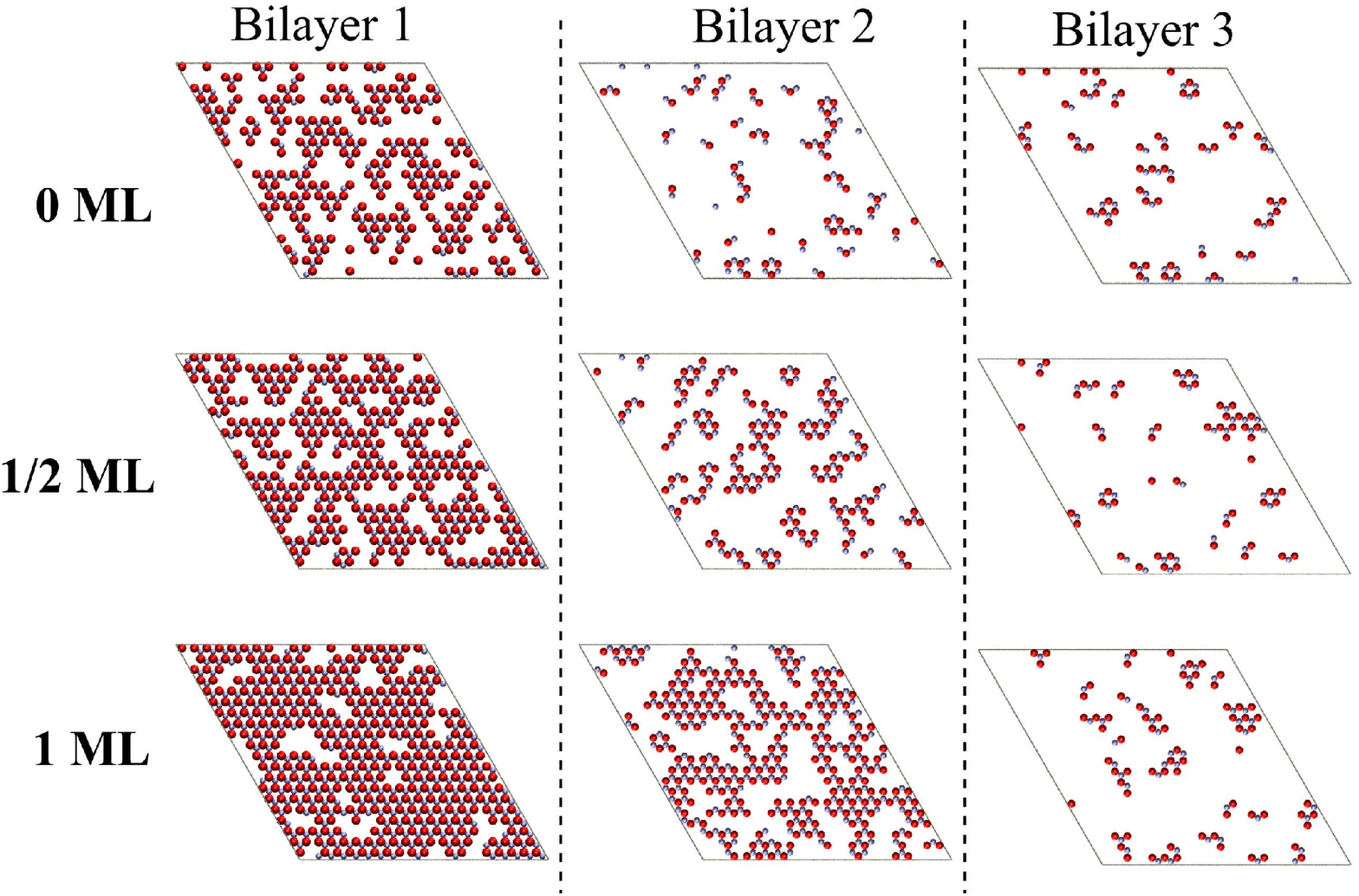} 

\caption{In a way similar to \ref{fig:04-Snapshot-of-three bilayers} we show snapshots of the three upper bilayers as 
the deposition of a single monolayer (ML) is carried out. The top row shows the three bilayers before the deposition is 
started, the middle one after 0.5 ML was deposited, and the lower one after a full ML was deposited onto the surface. 
Note that the triangular features of the reconstruction are essentially conserved throughout the 
deposition.\label{fig:06-deposition snapshots}}
\par
\end{centering}

\end{figure*}

The quenched MC simulation clearly shows that the system can evolve to a structure with preferred triangular motif and 
no regular tessellation of the surface. It is of interest to see if this effect also emerges during a simulation where 
the total number of ions in the surface layers increases over time, mimicking epitaxial growth. Starting from the end 
of a quenched simulation, we perform the MC annealling but now add pairs of Zn$^{2+}$/O$^{2-}$ ions at separate, 
randomly chosen sites in the three uppermost layers of the slab. Results from a simulation performed on a $N_{z}=6$, 
$N_{s}=32$ slab ($N=8,800$ ) are shown in Fig. \ref{fig:06-deposition snapshots}. Again, the simulation shows that 
triangular reconstructions form spontaneously, and grow larger by nucleation from characteristic smaller aggregates. We 
also see that the growth on the 2nd bilayer proceeds before the 1st bilayer is complete, leading to a surface with 
multiple ad-islands and pits of various size.

In Fig. \ref{fig:05-surface-energy_dipole} we also show how the surface energy and dipole moment change during the 
deposition simulation. The addition of the ions allows the system to find structures with decreasing magnitude of 
dipole moment, since thereby lowering the electrostatic energy. An interesting feature shown by Fig. 
\ref{fig:05-surface-energy_dipole} is that both surface energy and dipole moment oscillate as more ions are added, and 
reach even lower values as we deposit more and more ions. Thus the system finds a steady state with lower energy and 
dipole moment by making use of the larger number of degrees of freedom made available by the deposited ions.

\section{Discussion and Conclusion}

It is observed from the STM results \cite{dulub_novel_2003} that not one but several triangular features co-exist at 
the surface of Zn-terminated ZnO(0001). This behaviour also emerges in our Monte-Carlo simulations, for both the 
quenched structure in Fig. \ref{fig:04-Snapshot-of-three bilayers} and the deposition structure in Fig. 
\ref{fig:06-deposition snapshots}. The reason for this can be traced to the small energy differences between the 
various surface reconstructions shown in Table \ref{tab:surface-energy}; there is no substantially preferred reconstruction, provided 
$\theta_{\mbox{vac}}$ is close to 0.235 locally. Therefore the patterns that emerge in the MC simulations result from 
the competitive growth of energetically comparable triangular reconstructions. For this reason, we do not observe any 
long time coarsening of the structures in the simulation, as can be seen from the surface energy evolution shown in 
Fig. \ref{fig:05-surface-energy_dipole} during the quenched phase before deposition starts, and neither do we observe 
the system being restored to its regular tessellated starting configuration.

Our quenched MC simulation in Fig. \ref{fig:04-Snapshot-of-three bilayers} produces an interesting surface morphology 
that is reminiscent of the experimental STM images \cite{dulub_novel_2003,kresse_competing_2003}. Furthermore, the 
simulation with increasing surface coverage of Fig. \ref{fig:06-deposition snapshots} also includes other morphological 
features such as co-existing ad-islands and pits that are found experimentally. Therefore, we believe that the 
simulations capture some of the main physical processes that give rise to these surface reconstructions. However, it is 
important to note that the reconstructions found by the simulation are much smaller than those observed experimentally. 
The largest triangle observed from the MC simulation has a side of the order of 20\AA thus only reproducing the 
smallest clusters observed in the STM scans. It is possible that this is due to the lattice sizes and simulation 
durations that are accessible. Another explanation is that our use of formal charges ($\pm2e$) on the ions is not 
justified in this system. Using formal charges probably overestimates the strength of the Coulomb interaction, thereby 
tending to make the triangular clusters at the surface more compact. Work to augment the current model using a charge 
equilibration (or QEq) scheme \cite{rappe_jpc_1991} is therefore planned.

In conclusion, we believe that the models presented here do help to explain the Zn-terminated surface reconstructions 
observed experimentally. Whilst the accuracy of our models cannot compete with DFT, we are not restricted to studying 
individual reconstructions. The use of empirical potentials allows us to explore the phenomenology of the surface 
reconstructions more freely, and we find that a broad range of characteristic triangular motifs naturally emerge in our 
simulations, qualitatively consistent with the STM results.

Acknowledgement: This work was supported by the UK's Engineering and Physical Sciences Research Council grant 
EP/C524349 and by the University of Strathclyde.


\end{document}